\documentclass[referee, useAMS, usenatbib]{mn2e}
\usepackage{psfig}
\usepackage{graphicx}
\usepackage{epsf}
\usepackage{bm}
\usepackage{lscape}

\title [Identification of dwarfs and giants in RAVE]
{Identification of field dwarfs and giants in RAVE}

\author[Bilir et al.]
       {S. Bilir,${^1} \thanks{E-mail: sbilir@istanbul.edu.tr}$
        S. Karaali$^{2}$, S. Ak$^{1}$, \"O. \"Onal$^{1}$,
        B. Co\c skuno\u glu$^{1}$, G. M. Seabroke$^{3}$
\\
  $^1$Istanbul University Science Faculty, Department of Astronomy and Space
Sciences, 34119, University-Istanbul, Turkey\\
  $^2$Beykent University, Faculty of Science and Letters, Department of Mathematics
and Computer, Beykent 34398, Istanbul, Turkey\\
  $^3$Mullard Space Science Laboratory, University College London, Hombury St
      Mary, Dorking, RH5 6NT, UK\\}
\date{Accepted 2011 month day.
Received year month day; }

\pagerange{\pageref{firstpage}--\pageref{lastpage}} \pubyear{2007}

\begin{document}

\maketitle

\label{firstpage}

\begin{abstract}
The second RAdial Velocity Experiment (RAVE) Data Release (DR2) derives
$\log g$ values but we present a simpler and cleaner method of identifying
dwarfs and giants using only magnitudes, which does not require spectroscopic
analysis.  We confirm the \citet{Bilir06a} procedure  which estimates the number
of dwarfs and giants via their positions in the $J-V$ two magnitude diagram by
applying it to RAVE DR2. It is effective in estimating the number of dwarfs and
giants at $J-H>0.4$ compared to RAVE's $\log g$ values. For $J-H\leq0.4$, where
dwarfs and subgiants show a continuous transition in the $J$ magnitude histogram,
we used the Besan\c con Galaxy model predictions to statistically isolate giants.
The percentages of giants for red stars and for the whole sample are 85\% and
34\%, respectively. If we add the subgiants, the percentage of evolved stars for
the whole sample raises to 59\%. For the first time in the literature, we analysed
the effect of CHISQ on RAVE's $\log g$ values (CHISQ is the penalised $\chi^2$
from RAVE's technique of finding an optimal match between the observed spectrum
and synthetic spectra to derive stellar parameters). Neither the CHISQ values nor
the signal-to-noise ratio bias RAVE $\log g$ values.  Therefore the method of
identifying dwarfs and giants via the two magnitude diagram has been verified
against an unbiased dataset.
\end{abstract}


\section{Introduction}

Large surveys such as the Two Micron All Sky Survey \citep[2MASS,][]{2mass06},
Sloan Digital Sky Survey/Sloan Extension for Galactic Understanding and
Exploration \citep[SDSS/SEGUE,][]{York00, Yanny09} and the RAdial Velocity
Experiment \citep[RAVE,][]{Steinmetz06} give the opportunity to a researcher
to investigate different topics. However, one needs to select the appropriate
data from these surveys and to classify them according to his/her aim. RAVE
provides spectroscopic data observed from the southern hemisphere. Radial
velocities are available in the first data release \citep[DR1,][]{Steinmetz06}
and spectroscopic analyses to provide information on values of stellar
parameters (temperature, surface gravity and metallicity) are also included
in the second  data release \citep[DR2,][]{Zwitter08} additional to the
radial velocities. One can find the stellar parameters in the literature
supplemented by stellar position, proper motion and photometric measurements
from the Deep Near Infrared Survey of the Southern Sky \citep[DENIS,][]{Fouque00},
{\em 2MASS} \citep{2mass06} and Tycho-2 \citep{Hog00} surveys.

Examples of scientific use of such a dataset are described in \citet{Steinmetz03}.
They include the identification and study of the current structure of the
Galaxy and of remnants of its formation, recent accretion events, as well as
discovery of individual peculiar objects and spectroscopic binary stars.
For example, kinematic information derived from the RAVE dataset has been
used to demonstrate the presence of a dark halo in the Galaxy \citep{Smith07}.
 \citet{Veltz08} identified kinematical discontinuities in the direction to
the Galactic poles which separate a thin disc, thick disc and a hotter component.
\citet{Seabroke08} searched for infalling stellar streams onto the local
Galactic disk and found that it is devoid of any vertically coherent
streams containing hundreds of stars. \citet{Coskunoglu11} estimated
Local Standard of Rest (LSR) values from RAVE DR3.  \citet{Klement11} classified the
dwarfs and giants in RAVE and used this classification in stellar stream
detection.

Classification of stars in RAVE as dwarf and giant populations is also
the topic of this paper. However, the procedure we use here is different.
We use the positions of stars in the $J-V$ two magnitude diagram to
estimate the number of dwarfs and giants of the sample considered.
According to this procedure, stars which lie above the line fixed by
\citet{Bilir06a} are dwarfs and those below the same line are giants.
We show that this procedure provides accurate estimations compared to RAVE DR2
$\log g$ values. Also, it can be applied without the following considerations
of \citet{Klement11} and hence provides simplicity: {\em i}) the gravity of a
given luminosity class changes with colour, {\em ii}) reddening of stars causes
to a less clear separation of dwarfs from other luminosity classes in $\log g$
distribution, {\em iii}) the dwarf-giant separation gets better with increasing
colour, {\em iv}) the expected small fraction of dwarfs in combination with
their higher $\log g$ uncertainties might hinder a clear separation and
{\em v}) the $\log g$ distribution has been shown to vary with Galactic latitude.

Our work is the first one in the literature to analyse the effect of CHISQ in the
RAVE data. CHISQ is the penalised $\chi^2$ from RAVE's technique of
finding an optimal match between the observed spectrum and synthetic spectra to
derive stellar parameters. It is conceivable that sample stars with large CHISQ
and/or small signal-to-noise ratios (S/N) contaminate the dwarf
and giant separation with unreliable $\log g$ values. We investigate whether
CHISQ or S/N biases RAVE $\log g$ values and therefore biases the verification
of our method of identifying dwarfs and giants via the two magnitude diagram.

The paper is organized as follows: the data and field
dwarf-giant separation are given in Sections 2 and 3, respectively.
Confirmation of field dwarf-giant separation and the effects of S/N and
$\chi^2$ on the separation of dwarfs and giants are presented in Sections
4 and 5 and finally a summary and discussion is devoted to Section 6.

\section{Data}
The data were taken from the RAVE DR2 \citep{Zwitter08}. Among 51~829
objects contained in RAVE DR2, only 49~327 objects are unique, the
rest are repeated observations of single stars \citep{Zwitter08}.
We prepared our sample by applying the following constraints: {\em i})
$0<\log g\leq5$ and {\em ii}) $7.7 < V_{T}\leq12.6$, where $V_{T}$
is the  visual apparent magnitude in Tycho-2 \citep{Hog00}. Thus,
the total number of our sample was reduced to 11~470 stars, from which
we expect more reliable surface gravities and magnitudes. Along with
$\log g$ values, {\em 2MASS} ($J$, $H$ and $K_{s}$) and Tycho-2
($B_{T}$, $V_{T}$) apparent magnitudes are also available in RAVE
DR2. As we need Johnson's apparent $V$ magnitude, we transformed
the magnitudes $B_{T}$ and $V_{T}$ to Johnson's $V$ magnitude by
the following equation in the Hipparcos catalogue \citep{ESA97}:

\begin{equation}
V=V_{T}�-0.09\times(B_{T}�-V_{T})
\end{equation}
The errors in Tycho magnitudes propagate 0.16 mag for $V$ magnitude.

\citet{Zwitter10} estimates the distances of K0 dwarfs to be between 50
and 250 pc, which encouraged us to neglect extinction for dwarfs.
Although K0 giants are located at distances 0.7-3 kpc, their
Galactic latitudes are absolutely larger than 20$^\circ$. Hence,
no dereddening was applied to the $V$ and $J$ apparent magnitudes
used in the procedure which separates dwarfs and giants in our work.

To confirm our argument, we adopted the $(J-H)=0.27$ and $(J-H)=0.52$ (see Fig. 1) as the typical colours
of FG dwarfs and K giants, respectively and transformed them to $M_J$
magnitudes, i.e. 3.35 and 1.67 via the calibrations of \citet{Bilir08}
and the table of \citet{Covey07}. The combination
of these absolute magnitudes with the apparent $J$ magnitudes
$5.5\leq J \leq12$ provides distance for a given star. We estimated
$d~\sin(b)$ distances to the Galactic plane for a set of directions
defined by the Galactic latitudes $b=(30^{\degr}, 50^{\degr}, 75^{\degr})$
and Galactic longitudes $l=(0^{\degr},90^{\degr}, 180^{\degr},270^{\degr})$
and transformed the corresponding extinctions from \citet{Schlegel98} maps
to the  distance in question by the following equation \citep{Bahcall80}:

\begin{equation}
A_{d}(b)=A_{\infty}(b)\Biggl[1-\exp\Biggl(\frac{-\mid
d~\sin(b)\mid}{H}\Biggr)\Biggr].
\end{equation}
Here, $b$ and $d$ are the Galactic latitude and distance of the
star, respectively. $H$ is the scaleheight for the interstellar
dust which is adopted as 125 pc \citep{Marshall06} and
$A_{\infty}(b)$ and $A_{d}(b)$ are the total absorptions for the
model and for the distance to the star, respectively.
$A _{\infty}(b)$ can be evaluated by means of Eq. (3):

\begin{equation}
A_{\infty}(b)=3.1E_{\infty}(B-V).
\end{equation}
$E_{\infty}(B-V)$ is the colour excess for the model taken from the
NASA Extragalactic
Database\footnote{http://nedwww.ipac.caltech.edu/forms/calculator.html}.
Then, $E_{d}(B-V)$, i.e. the colour excess for the corresponding
star at the distance $d$, can be evaluated by Eq. (4) adopted for
distance $d$,

\begin{equation}
E_{d}(B-V)=A_{d}(b)~/~3.1.
\end{equation}
The numerical ranges for $E_{d}(B-V)$ for dwarfs
and giants, evaluated by this procedure for $|b|>20^{\degr}$, are [0.001,
0.063], [0.003, 0.071], respectively. Hence, we can neglect the interstellar
extinction in our work.

\section{Identification of dwarfs and giants via two bands}
We adopted the procedure of \citet{Bilir06a} to identify the
dwarfs and giants in our sample. \citet{Bilir06a} generated
an empirical separation line to define dwarf and giant stars in the two
selected star fields by their {\em 2MASS} and $V$ magnitudes. The
separation lines were formulated through $V-J$, $J-H$ and $V-K_{S}$
magnitudes diagrams in these equations: $J=0.957\times V-1.079$;
$H=0.931\times V-1.240$; $K_{s}=0.927\times V-1.292$. According
to this procedure, stars which lie above this line were classified
as dwarfs, whereas those lie below this line were classified as giants.
In this study, $J$ versus $V$ magnitude diagrams are used and stars with
$J>0.957\times V-1.079$ are expected to be dwarfs whereas those with
$J\leq0.957\times V-1.079$ are expected to be giants. The procedure
was derived from spectroscopic data, hence its confidence level is high.
We approach this problem as explained in the following.

\begin{figure*}
\center
\includegraphics[scale=0.50, angle=0]{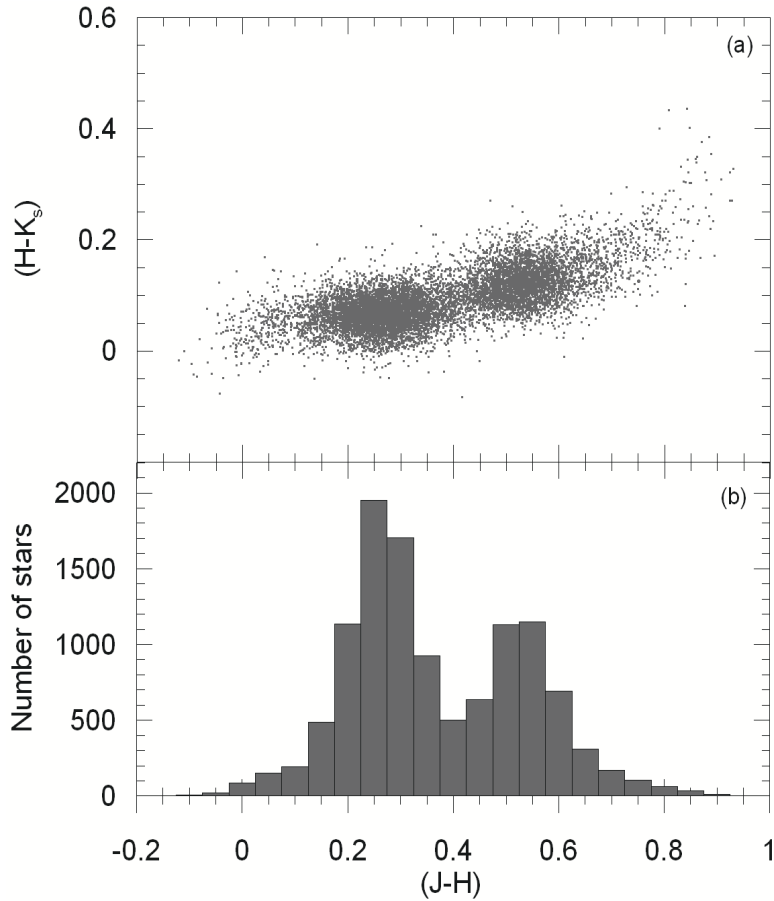}
\caption{The star sample divided into two subsamples at $J-H=0.4$
in the $(J-H)-(H-K_s)$ two-colour diagram (a). Distribution of
stars according to their $(J-H)$ colours (b).}
\end{figure*}

First of all, we plotted the $(J-H)$-$(H-K_{s})$ two colour diagram
of the sample stars. Fig. 1 reveals two subsamples separated by the
colour $J-H=0.4$ mag, keeping in mind that the dwarf-giant
separation is colour dependent. Dwarfs are characterized by larger
$\log g$ surface gravities relative to giants. Hence, we combined
$\log g$ with $J-H$ colours to proceed in our study. We adopted
$\log g=4$ as a rough limit for separating dwarfs and giants
(subgiants were not taken into account in the procedure of \citet{Bilir06a},
hence we do not consider them in this section, but see the following sections).
Thus  we applied the procedure of \citet{Bilir06a} to four subsamples of
stars, i.e. I: ($J-H\leq0.4$, $\log g\leq4$), II: ($J-H>0.4$,
$\log g\leq4$), III: ($J-H\leq0.4$, $\log g>4$) and IV: ($J-H>0.4$,
$\log g>4$), in order to separate dwarfs and giants in these
subsamples (see Table 1). Fig. 2 shows that the majority of stars (3567) in the
subsample II are giants, whereas only 484 stars which correspond
to 12\% of the total stars in this subsample are dwarfs. Stars in subsample III
consist of dwarfs (4357); only 1\% of the total stars in this subsample are giants
and the number of dwarfs and giants are almost equal in subsample IV.  In subsample I,
the number of dwarfs is 2520, corresponding to 99\% of the total number of stars in
this subsample.  This is an unexpected given that subsample I is defined to only
include stars with $\log g\leq 4$ i.e. the $\log g$ values should be too low to be dwarfs.

\begin{table}
\setlength{\tabcolsep}{4pt}
\centering
\caption{Comparison of percentages of dwarfs and giants in our RAVE sample.}
\begin{tabular}{ccccc}
\hline
& \multicolumn{2}{c}{Number of dwarfs} & \multicolumn{2}{c}{Number of giants} \\
\hline
Subsample & Total & \% & Total & \%  \\
\hline
$\log g\leq 4$, $J-H\leq0.4$ & 2520 & 98.78 &  31 &  1.22 \\
$\log g\leq 4$, $J-H>0.4$    &  484 &  11.95&   3567& 88.05\\
$\log g>4$, $J-H\leq0.4$ &  4357&  98.84 &   51&  1.16\\
$\log g>4$, $J-H>0.4$    &   211&   45.87&   249& 54.13\\
\hline
\end{tabular}
\end{table}

\begin{figure*}
\center
\includegraphics[scale=0.75, angle=0]{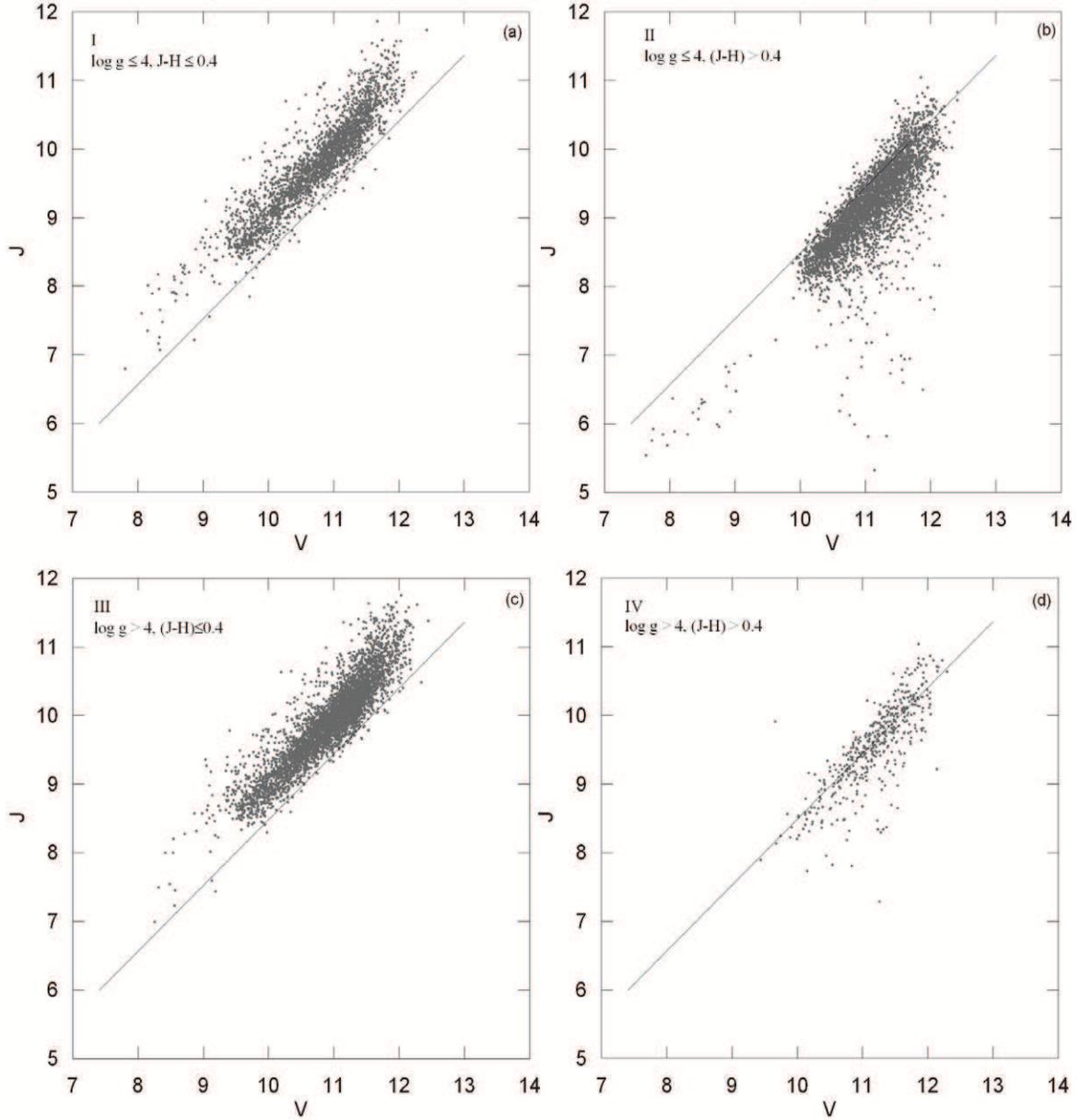}
\caption{The $J$ versus $V$ magnitude diagrams of four subsamples:
(a) I: $\log g\leq4$ and $J-H\leq 0.4$. (b) II: $\log g\leq4$ and
$J-H>0.4$. (c) III: $\log g>4$ and $J-H\leq 0.4$. (d) IV:
$\log g>4$ and $J-H>0.4$. The solid line represents the empirical
dwarf-giant separation line of \citet{Bilir06a}. The stars above
the line are dwarfs, whereas those below the line are giants.}
\end{figure*}

Hence, we separated the subsample I into three extra subsamples,
 i.e. Ia: $\log g\leq3$, Ib: $3<\log g\leq3.5$ and Ic:
$3.5<\log g\leq4$ (Fig. 3), in order to treat the problem in
detail. The  $J$ versus $V$  magnitude diagram for subsample
Ic classifies the majority of these stars as dwarfs: 2005 out of 2021 stars
(99\%), like subsample I. However, the number of dwarfs with
$3<\log g\leq3.5$ in subsample Ib and even those with
$\log g\leq3$ in Ia are still much larger than the number of giants,
i.e. they consist of more than 95\% of the total number of stars.
This is an unexpected result in the scheme of \citet{Bilir06a}, which
will be discussed in the following sections.

Additional to the error of 0.16 mag in $V$, mentioned in Section 2,
another error originates from the transformation given in \citet{Bilir06a}.
The errors 0.16 and 0.024 mag in $V$ and $J$, respectively, propagate
0.162 mag. This error causes a contamination of 12\% in the separation
of dwarfs and giants via $J-V$ two magnitude diagram. However, this contamination
is too small to be responsible for the results of subsamples I, Ia, Ib and Ic.

\section{Confirmation of the identification of dwarfs and
giants via two bands}
\subsection{Confirmation with the data appeared in the literature}

We confirm the identification of the dwarfs and giants in
our work by comparing their positions with \citet{Ammons06}'s
FGK dwarfs in the $V-J$ two magnitude diagram. \citet{Ammons06}
used spline functions of broadband photometry and estimated
fundamental astrophysical parameters, i.e. distance, effective
temperature, and metallicity, for more than 100~000 dwarfs. The
empirical broadband models are functions of Tycho-2 $(B_T,V_T)$
and {\em 2MASS} ($J$, $H$, $K_s$) magnitudes and proper motions for
FGK type stars.

We separated the stars of \citet{Ammons06} into six subsamples
defined by the $\log g$ surface gravities and $J-H$ colours
used for the definition of subsamples Ia, Ib, Ic, II, III
and IV (Table 2) and plotted them in Fig. 4. Afterwards,
we compared them with the stars investigated in our work.
We note that the stars of \citet{Ammons06}, which
were taken from the Tycho-2 catalogue, are partly observed in
the RAVE survey. The comparison of the two magnitude diagram positions of
the stars common to both \citet{Ammons06} and RAVE  is limited. However,
there are enough stars to obtain some results by this comparison (see Table 2).
The identification of dwarfs and giants by \citet{Ammons06} is based on
atmospheric parameters. Hence, its accuracy is high.

\begin{table}
\centering
\caption{Number of stars in six subsamples investigated
in our work and those which appear in \citet{Ammons06}.
The next two columns correspond to the number of stars
with signal to noise ratio less than 13 and 33,
respectively, and $\chi^2$ range.}
\begin{tabular}{cccccccccccccc}
\hline
& \multicolumn{5}{c}{\citet{Ammons06}} & \multicolumn{5}{c}{\citet{Bilir06a}} & \multicolumn{2}{c}{Signal-to-Noise} & \\
\hline
& \multicolumn{2}{c}{{Dwarf}} &\multicolumn{2}{c}{{Giant}}&    & \multicolumn{2}{c}{{Dwarf}}& \multicolumn{2}{c}{{Giant}}& & {S/N$\leq$13}& {S/N$\leq$33}&  \\
\hline
Subsample &  N & (\%)    & N &  (\%) & {Total} & N & (\%) & N & (\%)& {Total} & N & N & $\chi^2$ range\\
\hline
Ia & 48   & 98 & 2    & 4  & 50   & 92   & 95 & 5    & 5 & 97 & 2 & 31 & [240, 22502]\\
Ib & 276  & 98 & 5    & 2  & 281  & 423  & 98 & 10   & 2 & 433 & 7 & 86 & [074, 10579] \\
Ic & 1335 & 99 & 12   & 1  & 1347 & 2005 & 99 & 16   & 1 & 2021 & 13 & 329 & [075, 11826]\\
II & 118  & 04 & 3237 & 96 & 3355 & 484  & 12 & 3567 & 88 & 4051 & 18 & 428 & [024, 16633]\\
III& 2938 & 98 & 46   & 2  & 2984 & 4357 & 99 & 51   & 1 & 4408 & 47 & 950 & [046, 13029]\\
IV & 80   & 35 & 146  & 65 & 226  & 211  & 46 & 249  & 54 & 460 & 10 & 90 & [045, 09585]\\
\hline
Total     &4795& 100  &3448& 100  & 8243 &7572& 100  & 3898& 100& 11470 & 97 & 1914 \\
\hline
\end{tabular}
\end{table}

There are 48 dwarfs and two giants with $\log g\leq3$ and
$J-H\leq0.4$ in the catalogue of \citet{Ammons06} whose
positions fit with the dwarf and giant half planes in
Fig. 4a. Contrary to the expectation due to small surface
gravities, this is a strong confirmation that 92 stars
out of 97 in Fig. 3a are dwarfs.

\begin{figure}
\center
\includegraphics[scale=0.80, angle=0]{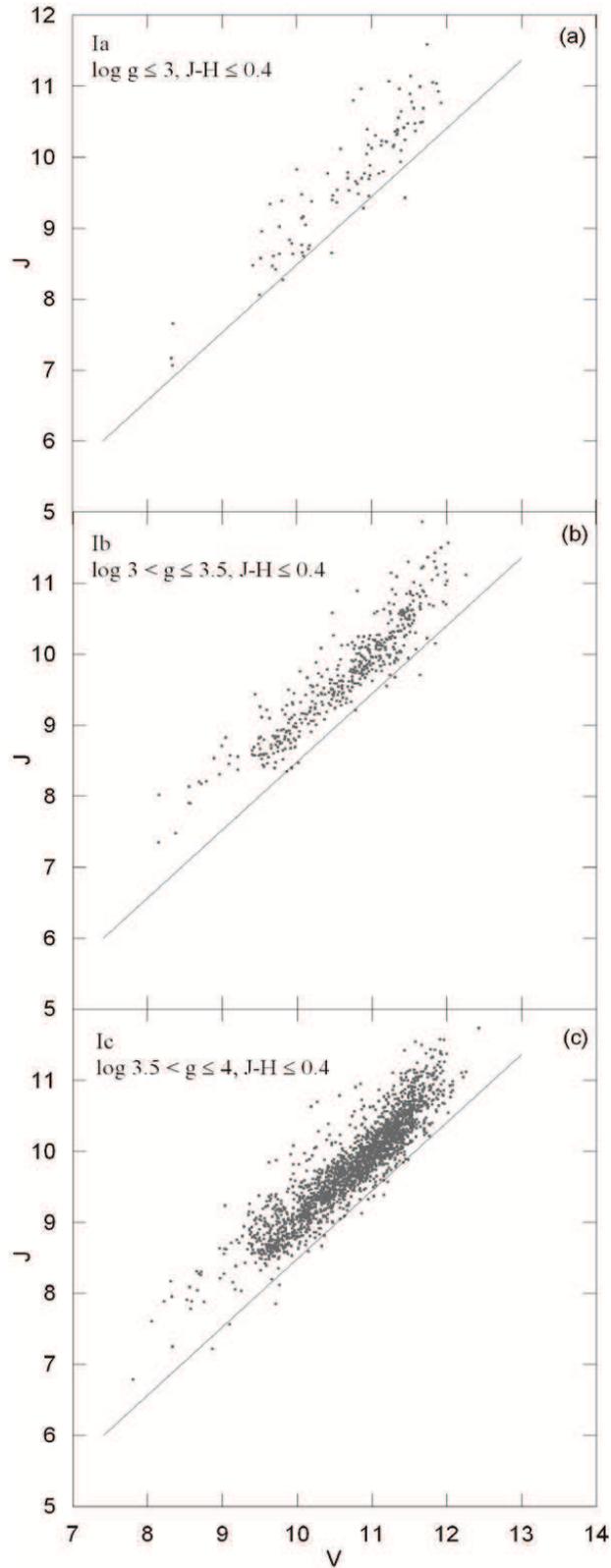}
\caption{Stars in subsample I ($\log g\leq4$ and $J-H\leq 0.4$)
divided into three extra subsamples according to their $\log g$
values: (a) Ia: $\log g\leq 3$. (b) Ib: $3<\log g\leq 3.5$.
(c) Ic: $3.5<\log g\leq 4$. The solid line represents the
empirical dwarf giant separation line of \citet{Bilir06a}.
The stars above the line are dwarfs, whereas those below the
line are giants.}
\end{figure}

\begin{figure*}
\center
\includegraphics[scale=0.80, angle=0]{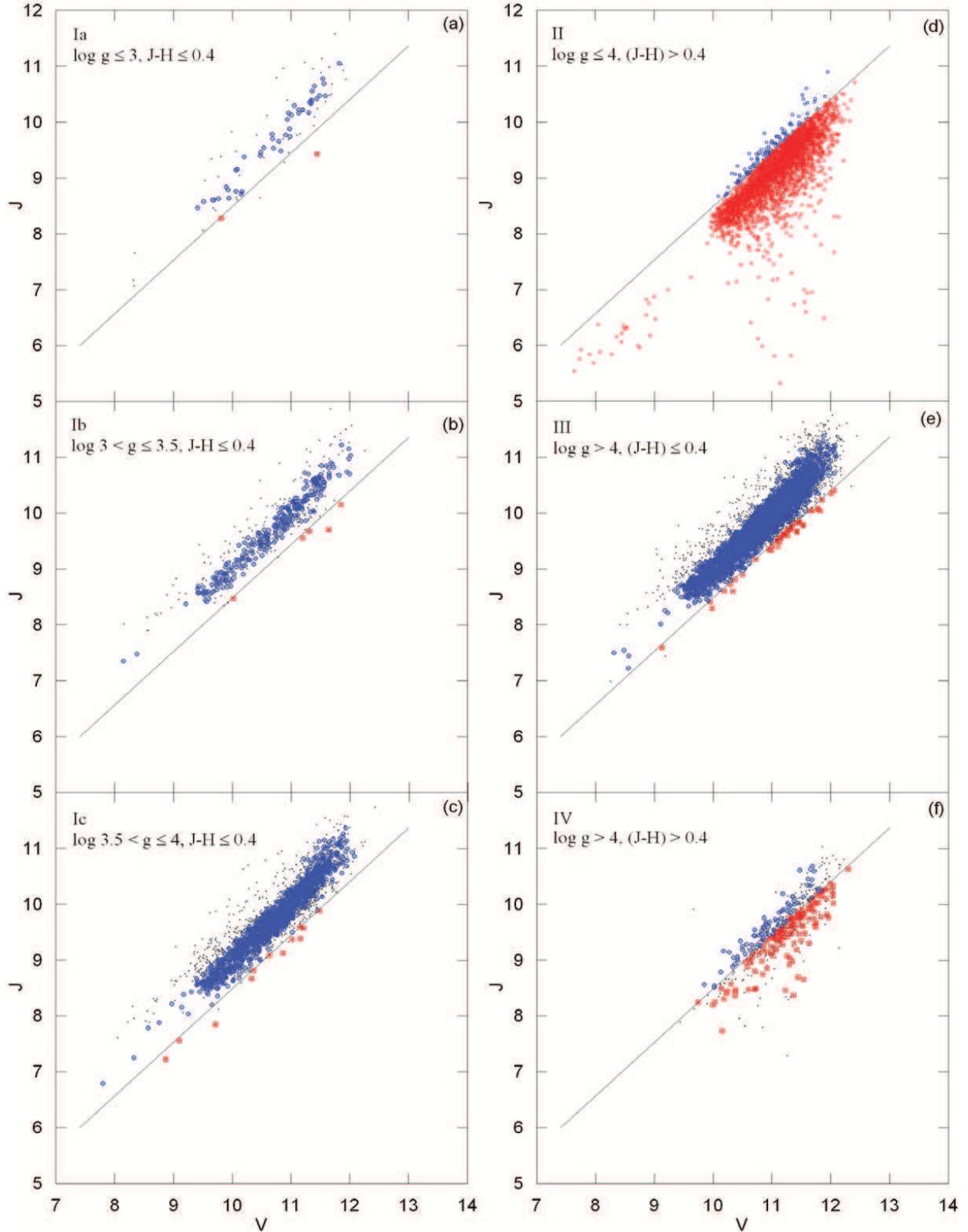}
\caption{Stars in six subsamples, i.e. (a) Ia, (b) Ib, (c) Ic,
(d) II, (e) III and (f) IV, plotted along with Ammons et al.
(2006)'s stars. Filled circles denote RAVE dwarfs and giants,
whereas open circles and squares denote dwarf and giant stars of
Ammons et al. (2006), respectively. The solid line represents
the empirical dwarf giant separation line of \citet{Bilir06a}.
The stars above the line are dwarfs, whereas those below the
line are giants.  The Roman numerals denote the subsamples
explained earlier.}
\end{figure*}

276 dwarfs and five giants of \citet{Ammons06} with $3<\log g\leq3.5$
and $J-H\leq0.4$ occupy the dwarf and giant half planes, respectively,
in Fig. 4b, confirming that 423 stars with the same surface gravities and
colours in our work which occupy the dwarf half plane are dwarfs. 10 stars
(2\% of the total stars) in Fig. 3b which occupy the giant half plane
should be giants.

We noted in the preceding section that most of the stars with $3.5<\log g\leq4$
and $J-H\leq0.4$ should be dwarfs. Actually, 2005 dwarfs, taken from
\citet{Ammons06}, with the same surface gravities and colours confirm our
suggestion (Fig. 4c). Additionally, 12 stars out of 16 (Fig. 3c) which were
classified as giants in \citet{Ammons06} occupy the giants' half plane,
confirming the procedure which separates dwarfs and giants via two bands.

The data of \citet{Ammons06} confirms also the results, obtained and noted
in Section 3, for subsamples II ($\log g\leq4$, $J-H>0.4$), III ($\log g>4$,
$J-H\leq0.4$) and IV ($\log g>4$, $J-H>0.4$) i.e. stars identified as dwarfs
and giants in \citet{Ammons06} occupy the dwarf and giant half planes in
the panels Fig. 4d, 4e and 4f, respectively. Giants dominate in panel Fig 4d,
whereas the majority of stars in panel Fig. 4e are dwarfs. As expected, panel
Fig. 4f involves both populations, giants and dwarfs, in large numbers.

The values in the two columns before the last one denote to the number
of stars with signal to noise ratio $(S/N)\leq33$ and $(S/N)\leq13$,
corresponding to the mean $(S/N)-1\sigma$ and $(S/N)-2\sigma$, respectively,
where $\sigma=20$ is the standard deviation of S/N. These stars do not show any
positional systematic difference relative to the ones with higher S/N, in the
$J-V$ two magnitude diagram. Finally, the values in the last column are the
CHISQ values (see Section 5 for details).

\subsection{Confirmation with surface gravity--colour diagram}
We plot all subsamples of stars, Ia, Ib, Ic, II, III and IV in the $\log g$
surface gravity and $J-H$ colour diagram and discuss their identifications
(Fig. 5). Furthermore, isochrones are overlaid on the data in order to distinguish
the dwarf region in the $J-H$ versus $\log g$ diagram. The isochrones are from the Padova database
\citep{Marigo08}, using the web interface\footnote {http://stev.oapd.inaf.it/cgi-bin/cmd}.
For our sample of stars, we adopted isochrones with metallicity $-2\leq[M/H]\leq+0.2$ dex
with 0.5 dex intervals for ages 1, 5 and 10 Gyrs. In total, 18 isochrones are plotted over
the data in the $\log g$ versus $J-H$ diagram (Fig. 5).

\begin{figure*}
\center
\includegraphics[scale=0.50, angle=0]{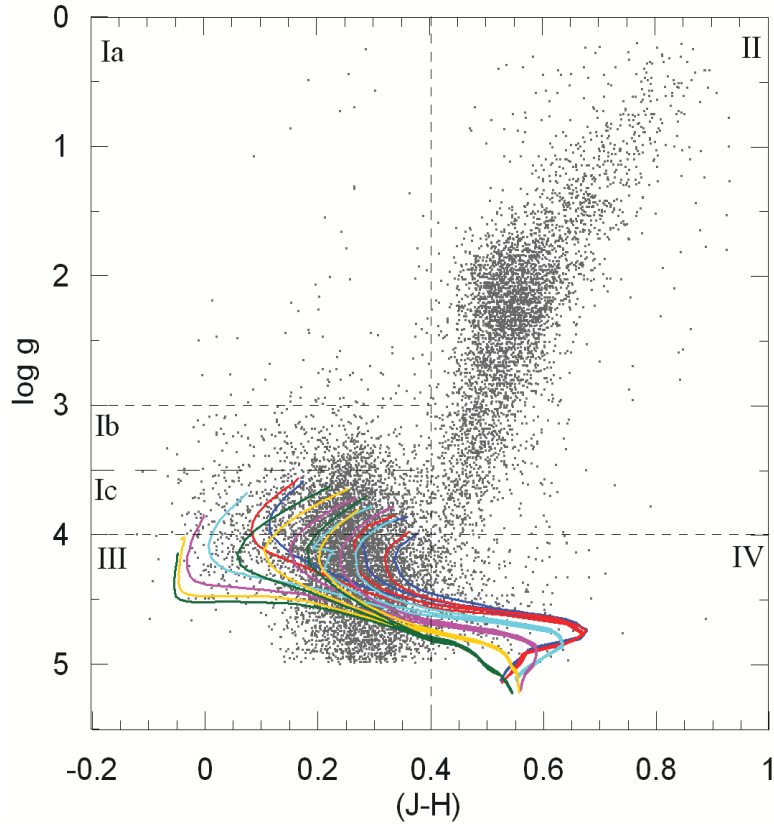}
\caption{The  $\log g$ versus $J-H$ diagram of the sample. Coloured lines denote
the Padova isochrones with different metallicities ($-2\leq[M/H]\leq0.2$ dex)
and ages (for 1, 5 and 10 Gyrs). The rectangles represent subsamples.}
\end{figure*}

Most of the stars in subsample II ($\log g\leq4$, $J-H>0.4$) should be
red giants. This is the case in Fig. 2b, i.e. 3567 giants and 484 dwarfs could be identified
according to their positions in the $J$ versus $V$ magnitude diagram. Dwarfs
correspond to only 12\% of the total subsample.

Another confirmation can be done easily for stars in subsample III ($\log g>4$,
$J�-H\leq0.4$). The large surface gravities and their positions on the disk
population isochrones of these stars indicate that they should be dwarfs.
Actually, the number of dwarfs identified by their position in Fig. 2c consists
of 99\% of the total stars of this subsample.

Almost half of the stars in subsample IV ($\log g>4$, $J-H>0.4$) lie on the
main sequence of the isochrones. Hence, these stars with large surface gravities
should be dwarfs.  The gap in the isochrones at $J-H>0.4$ between the dwarf branch
(shown in Fig. 5 panel IV) and the giant branch (giants are plotted in Fig. 5
panel II but the corresponding isochrone is not shown) suggests there should not
be any stars with $J-H>0.4$ and $4 < \log g < 4.5$.  However, Fig. 5 shows data
exist in this region.  These are most likely to be due to measurement errors on
$\log g$. RAVE DR2 $\log g$ errors are conservatively estimated to be 0.5 dex for
an average S/N $\sim$ 40. This is large enough that dwarfs with $\log g>4.5$ can
be measured to have $\log g>4$ and giants with $\log g<4$ can be measured to
have $\log g<4.5$. This agrees with our argument that the number of dwarfs and
giants in the subsample IV are almost equal (Fig. 2d).

Confirmation of dwarfs and giants identified by two bands, $J$ and $V$, for
subsamples Ia and Ib with the diagram surface gravity versus colour is not
easy (Fig. 5). Although 95\% of the total stars in subsample Ia ($\log g\leq3$,
$J-H\leq0.4$) occupy the dwarf half plane, their surface gravities are rather
small. Almost the same case holds for stars in  subsample Ib ($3<\log g\leq3.5$,
$J-H\leq0.4$) where surface gravities have an upper limit of $\log g=3.5$.
Most of the stars in the subsample Ic ($J-H\leq0.4$, $3.5<\log g\leq4$) lie on
the subgiants segments of the isochrones, indicating that they are subgiants
rather than dwarfs. Hence, we should add a procedure to the one of
\citet{Bilir06a} for separation of subgiants from dwarfs and giants.

\subsection{Number of Stars for Different Populations Estimated via Galaxy Model}
We use the Besan\c con Galaxy model \citep{Robin03} to estimate the number
of stars for different populations, i.e. dwarf, subgiant and giant, in
our sample and compared them with the ones obtained by using the procedure
of \citet{Bilir06a}. The Besan\c con Galaxy model of stellar population synthesis
is a simulation tool used to test Galaxy evolution scenarios by comparing
stellar distributions predicted by these scenarios with observations, such as
photometric star counts and kinematics. The model assumes that stars are
created from gas following a star formation history and an initial mass
function; stellar evolution follows evolutionary tracks \citep{Schultheis06}.

We applied the model to stars with $0<J-H\leq0.4$ (sample 1) and
$0.4<J-H\leq1$ (sample 2) separately. The restrictions used for the model
are as follows: $d\leq2$ kpc, size: one square degree, $7.7<V\leq12.6$, zero
absorption, $0<B-V\leq0.8$ (for sample 1) and $0.8<B-V\leq1.63$
(for sample 2). The model has been applied to 12 directions of the Galaxy
and the number of stars for each population has been estimated for each
direction. The sets for Galactic latitude and longitude combined
for the mentioned directions are: $|b|$=(30$^{o}$, 50$^{o}$, 70$^{o}$)
and $l$=(0$^{o}$, 90$^{o}$, 180$^{o}$, 270$^{o}$). The results are given
in Table 3. The number of stars for a particular population varies for
different Galactic directions, as expected. Hence, we adopted the mean
values for each population for sample 1 and sample 2.

As already seen in Fig. 5, there is an absence of subgiants in
the sample of stars with $0.4<J-H\leq1$ in Table 3,which gives us the chance to
compare the number of giants and dwarfs claimed in Table 2 with the
ones estimated via the Besan\c con Galaxy model. The number of giants
in subsample II and subsample IV is 3816 which corresponds to 85\% of
the total number of stars with $0.4<J-H\leq1$. This is rather close
to the corresponding model value, 82\% (see Table 3), and of course the same holds
for dwarfs, i.e. 15\% observed and 18\% in the model.

The Besan\c con Galaxy model predicts that 33\% of the stars with $0<J-H\leq0.4$
will be subgiants and 2\% will be giants. The number of giants for this
colour claimed in Table 2 is 1\%, in close agreement with the model. However,
the procedure which separates dwarfs and giants via
two bands has not been scaled for subgiant separation. Hence, we
should find another procedure to carry out this work.

We plotted the histogram for $J$ magnitudes for stars with $0<J-�H\leq0.4$
and fitted it to a Gaussian curve (Fig. 6). The unique modality of the
distribution shows that there is a continuous transition between dwarfs
and subgiants when one goes from bright $J$ magnitudes to the faint
ones. A similar claim can be found in \citet{Klement11} for $\log g$ of
early type stars. However, we can find the number of subgiants in our
work statistically, by adopting the model percentages given in the
preceding paragraphs. 65\% of the stars in a Gaussian distribution
corresponds to 0.94 standard deviation of that distribution. Hence,
stars with $0<J-�H\leq0.4$ and, $\overline{J}<-0.94\times s$ and
$\overline{J}>+0.94\times s$ will be considered as evolved (subgiants
and giants) stars. Here, $\overline{J}$ and $s$ correspond to the mean
and standard deviation, respectively. Thus, the total number of evolved
stars and dwarfs in question are 2900 and 4060, respectively.

\begin{table}
\setlength{\tabcolsep}{4pt}
\centering
\caption{Percentages of dwarfs, subgiants and giants for our two $J-H$ colour
intervals, for different Galactic directions estimated via the Besan\c con
Galaxy model. The symbols  $l$ and $b$ are Galactic longitude and latitude respectively.}
\begin{tabular}{rcccccccc}
\hline
 \multicolumn{3}{r}{Colour range $\rightarrow$}&    \multicolumn{3}{c}{$0<J-H\leq0.4$} &     \multicolumn{3}{c}{$0.4<J-H\leq1$}\\
\hline
{Galatic latitude $(^{o})$} &    {$l(^{o})$} &    {$b(^{o})$} & Giant (\%) & Subgiant (\%) & Dwarf (\%) & Giant (\%) & Subgiant (\%) & Dwarf (\%)\\
\hline
$20<|b|<40$&          0 &         30 &       5.88 &      43.14 &      50.98 &      90.91 &       0.00 &       9.09 \\
           &         90 &         30 &       5.26 &      45.61 &      49.12 &      88.24 &       0.00 &      11.76 \\
           &        180 &         30 &       2.22 &      42.22 &      55.56 &      84.62 &       0.00 &      15.38 \\
           &        270 &         30 &       6.35 &      39.68 &      53.97 &      92.00 &       0.00 &       8.00 \\
$40<|b|<60$&          0 &         50 &       2.86 &      31.43 &      65.71 &      80.00 &       0.00 &      20.00 \\
           &         90 &         50 &       0.00 &      33.33 &      66.67 &      93.33 &       0.00 &       6.67 \\
           &        180 &         50 &       0.00 &      53.57 &      46.43 &     100.00 &       0.00 &       0.00 \\
           &        270 &         50 &       0.00 &      19.35 &      80.65 &      75.00 &       0.00 &      25.00 \\
$60<|b|<90$&          0 &         75 &       0.00 &      20.00 &      80.00 &      66.67 &       0.00 &      33.33 \\
           &         90 &         75 &       0.00 &       8.33 &      91.67 &      60.00 &       0.00 &      40.00 \\
           &        180 &         75 &       3.57 &      25.00 &      71.43 &      83.33 &       0.00 &      16.67 \\
           &        270 &         75 &       0.00 &      33.33 &      66.67 &      66.67 &       0.00 &      33.33 \\
\hline
           & \multicolumn{2}{r}{Average (\%)}&    2.18 &      32.92 &      64.91 &      81.73 &       0.00 &      18.27 \\
           & \multicolumn{2}{r}{This paper (\%)}& 1.18 &\multicolumn{2}{c}{98.82$^{*}$}& 84.59 &\multicolumn{2}{c}{15.41$^{*}$}\\
\hline
\multicolumn{9}{l}{$*$ This symbol shows the percentage of the total number of subgiant and dwarf stars.}\\
\end{tabular}
\end{table}

\begin{figure}
\center
\includegraphics[scale=0.50, angle=0]{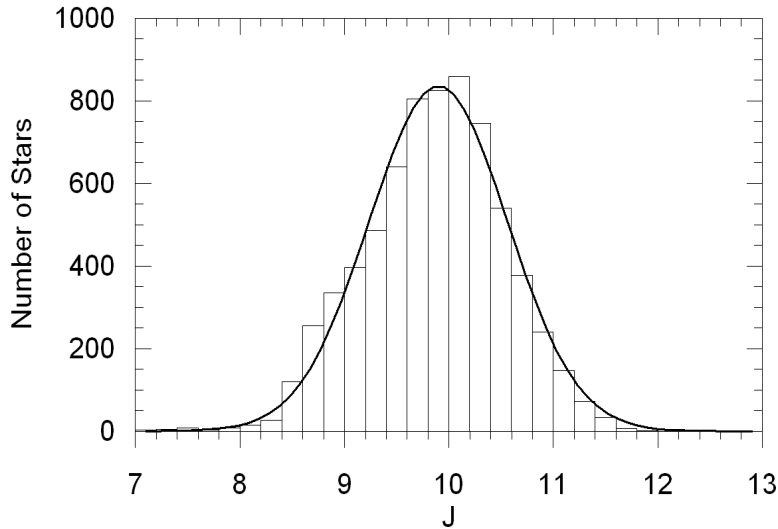}
\caption{Histogram for$J$ magnitudes for stars with $0<J-H\leq0.4$ fitted
to a Gaussian curve. Single mode indicates to the continuous transition between
dwarfs and subgiants (see the text).}
\end{figure}

\section{Effect of S/N ratio and CHISQ Values on identification of dwarfs and giants}
\subsection{The effect of S/N}
Although the identification of dwarfs and giants via two bands have been
confirmed in Section 4, one can argue that the positions of some stars in our
work may be erroneously determined. Such an argument should be discussed for at
least some subsamples of stars, such as subsample I ($\log g\leq4$, $J-H\leq0.4$).
In subsample I, the majority of the involved stars are identified as dwarfs,
 though their surface gravities are small. If such a contamination exists it
may originate from the small S/N ratios. For example as S/N decreases, the
wings of the spectral lines become more affected by noise, making them appear
narrower, which could mimic a lower $\log g$.

The mean and the median of S/N ratios for all stars in our sample (11470 stars)
 are close to each other which indicates that S/N ratios form a Gaussian
distribution. Their mean and standard deviation are 53 and 20, respectively. We
adopted the 1914 stars with S/N ratios less than 33 (53-20) as candidates of
contaminating our diagrams, and  marked them with blue colours on the $J$ versus
$V$ magnitude diagrams (Fig. 7) to treat the effect of S/N ratios on the
identification of dwarfs and giants. Stars with $S/N>33$ ratios are plotted
on the same diagrams with their original symbols. One can not reveal any
systematic scattering between two sets of S/N ratios, for subsamples I, II,
III, and IV defined in the previous sections. For example, the S/N ratios for
dwarfs with small $\log g$ surface gravities may be larger than 33 as well as
smaller than this numerical value. The same result has been obtained for
mean (S/N)-$2\sigma$ standard deviations, i.e. $S/N\leq13$. Our conclusion
is that small S/N ratios do not affect the identification of dwarfs and
giants via two bands. The percentages of the stars with $S/N\leq33$
are given in Table 4.

\begin{figure}
\center
\includegraphics[scale=0.75, angle=0]{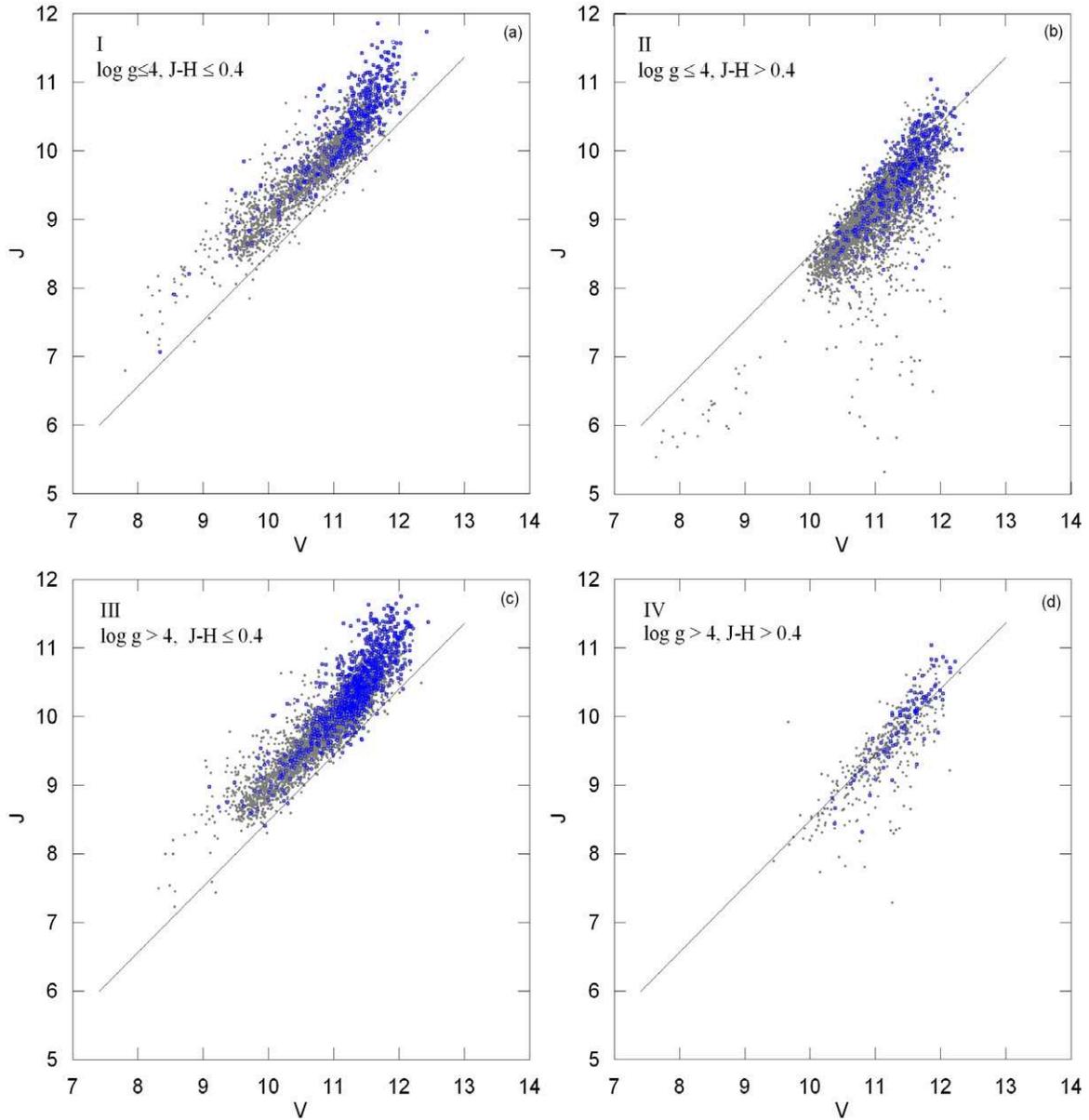}
\caption{Investigating the effect of S/N ratios on the identification of dwarfs
and giants. Stars with $S/N\leq 33$ (open blue circles) do not show any systematic
distribution relative to stars which are confirmed as being giants/dwarfs
(filled black circles).}
\end{figure}

\subsection{The effect of CHISQ}
We checked whether the CHISQ values in RAVE DR2 bias the identification
of dwarfs and giants. CHISQ is the penalised chi-squared from the technique
finding an optimal match between the observed spectrum and the one constructed
from a library of pre�computed synthetic spectra to derive stellar
parameters, including $\log g$. The spectral type of RAVE stars is generally
not known and the input catalogue does not use any colour criterion. So,
RAVE stars are expected to include all evolutionary stages and a wide range
of masses in the HR diagram. However, the template library only covers
normal stars. So, peculiar objects cannot be classified correctly. Sometimes
a peculiar nature of the spectrum can be inferred from a poor match of the
templates, despite a high S/N of the observed spectrum. This poor match
of a spectrum to a template is quantified by CHISQ.

\begin{figure}
\center
\includegraphics[scale=0.45, angle=0]{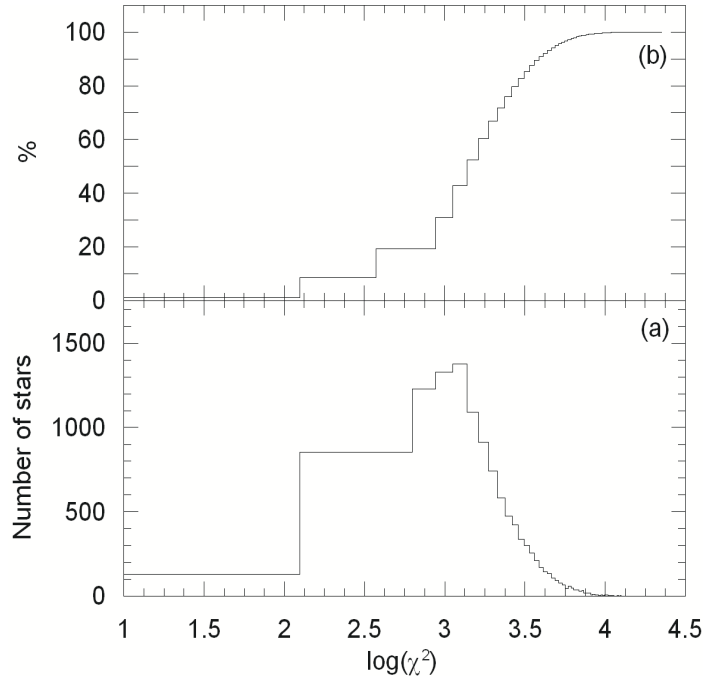}
\caption{CHISQ histogram for stars in our sample (a). Stars with CHISQ values larger
than 2290 consist 20\% of the sample (b).}
\end{figure}

We used the CHISQ values of stars in our sample to treat this problem.
These values lie between 24 and 22~503. We plotted their histogram in Fig. 8,
and colour-coded stars whose CHISQs are larger than 2290 (20\% of the whole
sample) in the $J$ versus $V$ two magnitude diagram (Fig. 9). The
distribution of CHISQ is not plotted or discussed in \citet{Zwitter08}. 
This is the first analysis of RAVE DR2's CHISQ values to appear in the
literature. As in Fig. 7, one can not reveal any systematic scattering
between stars with CHISQ values larger and smaller than 2290 (20\%). Hence, our
conclusion is that the CHISQ values in RAVE DR2 do not bias the
identification of dwarfs and giants via two bands. CHISQ distribution for
the sample stars is given in Table 5.

\begin{figure}
\center
\includegraphics[scale=0.75, angle=0]{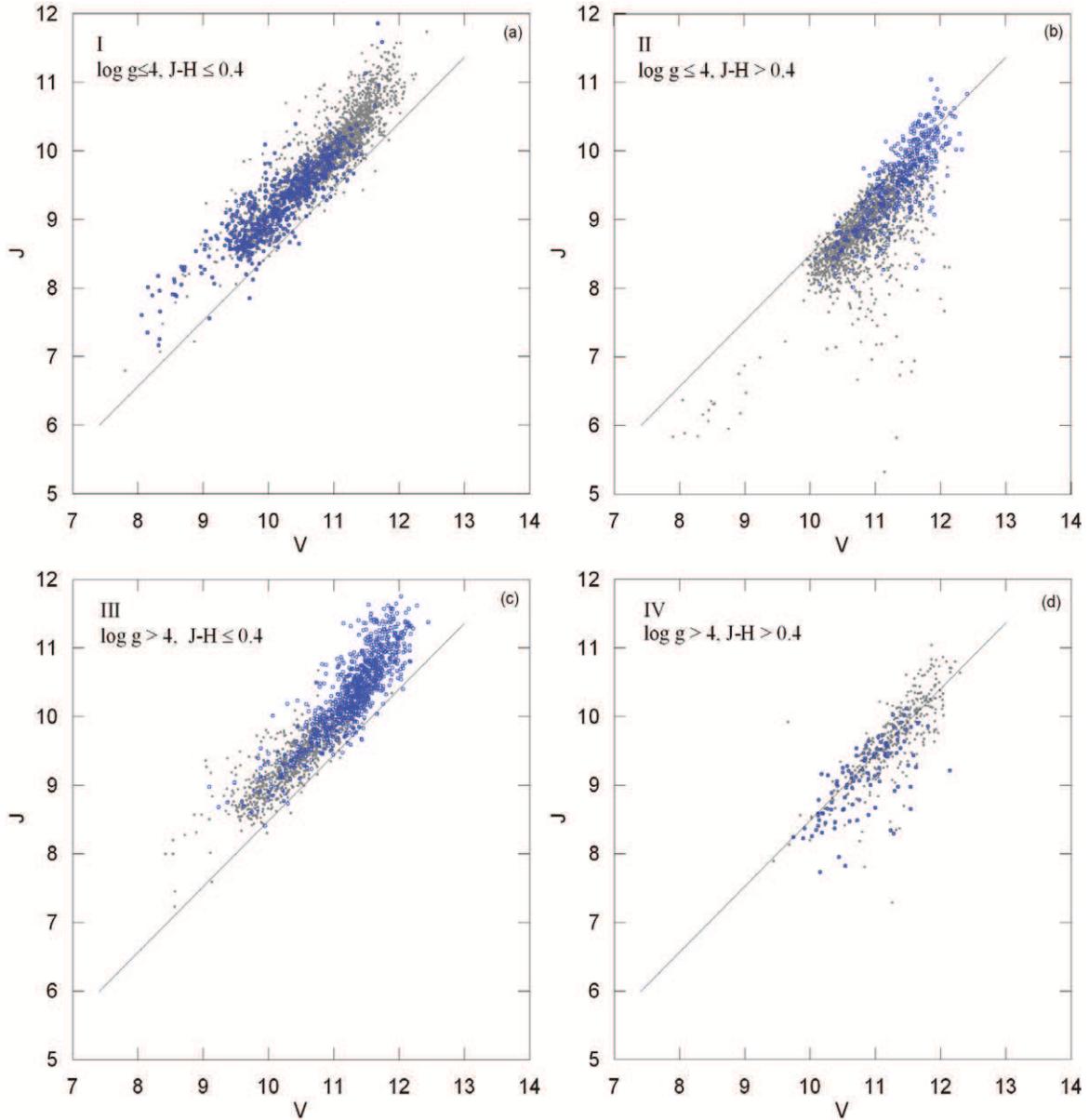}
\caption{Investigating the effect of CHISQ on the identification of dwarfs
and giants. Stars with CHISQ between 2290 and 22503 (20\% of the sample,
blue open circles) do not show any systematic distribution relative to
stars with CHISQ less than 2290 (filled black circles).}
\end{figure}

\section{Summary and Discussion}
We estimated the number of stars for different populations in the RAVE DR2 by
means of their positions in the $J$ versus $V$ two magnitude diagram
\citep{Bilir06a} and the Besan\c con Galaxy model \citep{Robin03}. The procedure
of \citet{Bilir06a} is scaled for separation of dwarfs and giants (but not
for subgiants). One can predict the number of subgiants, additional
to dwarfs and giants, via the Besan\c con Galaxy model. Different results were
obtained for two samples of stars, i.e. for blue and red ones, separated
by $J-H=0.4$.

{\bf a) Sample of stars with $J-H>0.4$:} Dwarfs and giants in this sample have been
separated in a very simple manner, i.e. by their positions in the $J-V$
two magnitude diagram. No additional constraints are needed for this separation.
The percentages of dwarfs (15\%) and giants (85\%) are confirmed by the percentages
of stars predicted by the Besan\c con Galaxy model, i.e. 18\% dwarfs, 82\% giants and
no subgiants. A second confirmation was done using the work of \citet{Ammons06}.
The positions of dwarf and giant in \citet{Ammons06} occupy the dwarf and giant
half planes in our work. Finally, the positions of the stars with $J-H>0.4$ in the
$\log g$ versus $J-H$ diagram indicate a large number of giants. Actually, most
of these red stars have surface gravities less than the upper limit attributed
to giants, i.e. $\log g=3.8$ \citep{Zwitter10}.

{\bf b) Sample of stars with $J-H\leq0.4$:} The separation of stars with
$J-H\leq0.4$ by means of their positions in the $J$ versus $V$ two magnitude
diagram is not as easy as for stars with $J-H>0.4$, because there is a
considerable number of subgiants among these stars which were not scaled
in \citet{Bilir06a}. The procedure of \citet{Bilir06a} reveals 82 giants and
6877 dwarfs corresponding to 1\% and 99\% of the total number of stars,
respectively. However, the percentage number of giants, dwarfs and subgiants
predicted by the Besan\c con model \citep{Robin03} are 2\%, 65\% and 33\%,
respectively. Hence, we plotted the histogram for $J$ magnitude, for stars
with $J-H\leq0.4$ to investigate this problem. The unique modality of the
distribution indicates that there is a continuous transition between dwarfs
and subgiants. Hence, we applied the Besan\c con Galaxy model to this sample
for estimation of the number stars for different populations. Thus, the
number of dwarfs reduced to 58\%, and the remaining 42\% of the total sample
turned out to be evolved stars (giants and subgiants).

It should be noted that dwarfs and giants with $J-H\leq0.4$ in
\citet{Ammons06} confirm the dwarf and giant half planes, respectively,
defined by $J$ and $V$ magnitudes. The identification of dwarfs and giants
by \citet{Ammons06} is based on atmospheric parameters. Hence, its accuracy
is high.

{\bf c) Error in $\log g$:} 18\% of the total stars with $3.5<\log g<4$
in Fig. 5 occupy the subgiant segments of the isochrones. Hence, they should
be subgiants. However, the  percentages of subgiants predicted by the
Besan\c con model is 25\%. The difference between percentages can be
explained by the estimated error on RAVE DR2 surface gravities being
as large as 0.5 dex.

{\bf d) Effect of S/N and CHISQ values:} We showed in this work that
neither the small signal to noise ratio (S/N) nor the CHISQ values
bias RAVE $\log g$ values. Therefore the method of identifying dwarfs
and giants via the two magnitude diagram has been verified against an
unbiased dataset.

{\bf e) Comparison between the results in our work and the ones apparent in
the literature:} The first estimate of the percentage of different stellar
populations in RAVE was published in \citet{Seabroke08}. They used a reduced
proper motion diagram to kinematically separate red ($J-K>0.5$) giants from
red ($J-K>0.5$) dwarfs. Statistically, they found that 44\% of their RAVE
sample consisted of red ($J-K>0.5$) giants and red ($J-K>0.5$) subgiants.
Although \citet{Seabroke08} do not explicitly state their 44\% sample
includes subgiants, their fig. 13 shows the subgiant branch extends to
redder colours than $J-K=0.5$. \citet{Seabroke08} do not distinguish between
blue ($J-K<0.5$) subgiants and blue ($J-K<0.5$) dwarfs. Therefore
 \citet{Seabroke08} do not provide an estimate for the percentage of
evolved (subgiant and giant) stars in RAVE. Given their 44\% sample excludes
blue ($J-K<0.5$) subgiants, 44\% represents a lower limit on the percentage
of evolved stars in RAVE, which is consist with the value of 59\% that we
find in our work. It should also be noted that the \citet{Seabroke08} RAVE
sample consists of many more stars than DR2 ($\sim$200~000 stars compared to
$\sim$50~000) so the density of sampling in different directions on the sky
will be different between the two samples and so they are not suitable for
detailed direct comparison. \citet{Klement11} found 43\% of RAVE DR2 are
dwarfs (see completeness column in their table 3). The remaining 57\% are
subgiants, giants and supergiants, which agrees closely with our value.

\begin{table}
\setlength{\tabcolsep}{4pt}
\centering
\caption{Comparison of percentages of dwarfs and giants with different ranges of S/N.}
\begin{tabular}{ccccccccccc}
\hline
& \multicolumn{5}{c}{Number of dwarfs} & \multicolumn{5}{c}{Number of giants}\\
\hline
Subsample & Total&  S/N$\leq$33 & \% & S/N$>$33 & \% & Total&  S/N$\leq$33 & \% & S/N$>$33 & \% \\
\hline
$\log g\leq 4$, $J-H\leq0.4$ & 2520 & 444 &  18 & 2076 & 82 & 31   & 2   & 7  & 29   & 93\\
$\log g\leq 4$, $J-H>0.4$    & 484  & 112 &  23 & 372  & 77 & 3567 & 316 & 9  & 3251 & 91\\
$\log g>4$, $J-H\leq0.4$     & 4357 & 939 &  22 & 3418 & 78 & 51   & 11  & 22 & 40   & 78\\
$\log g>4$, $J-H>0.4$        & 211  & 51  &  24 & 160  & 76 & 249  & 39  & 16 & 210  & 84\\
\hline
\end{tabular}
\end{table}

\begin{table}
\setlength{\tabcolsep}{4pt}
\centering
\caption{Comparison of percentages of dwarfs and giants with different ranges of CHISQ ($\chi^2$).}
\begin{tabular}{ccccccccccc}
\hline
& \multicolumn{5}{c}{Number of dwarfs} & \multicolumn{5}{c}{Number of giants}\\
\hline
Subsample & Total&  $\chi^2$$\leq$2290& \% & $\chi^2$$>$2290& \% & Total&  $\chi^2$$\leq$2290& \% & $\chi^2$$>$2290& \% \\
\hline
$\log g\leq 4$, $J-H\leq0.4$ & 2520 & 1875 & 74 & 645 & 26 &   31 &   16 & 52 &   15& 48\\
$\log g\leq 4$, $J-H>0.4$    &  484 &  414 & 86 &  70 & 14 & 3567 & 2083 & 58 & 1484& 42\\
$\log g>4$, $J-H\leq0.4$     & 4357 & 3541 & 81 & 815 & 19 &   51 &   40 & 78 &   11& 22\\
$\log g>4$, $J-H>0.4$        &  211 &  177 & 84 &  34 & 16 &  249 &  178 & 71 &   71& 39\\
\hline
\end{tabular}
\end{table}

As conclusion, stars with $J-H>0.4$ can be separated into dwarf and giant populations
in a very simple manner, i.e. by their positions in the $J-H$ two magnitude
diagram. One does not need any other constraints for such a separation. Our argument
has been confirmed by the work of \citet{Ammons06} and the Besan\c con Galaxy
model \citep{Robin03}. However, the existence of a considerable number of
subgiants complicates the separation of stars with $J-H\leq0.4$ into different
population types. The difference between the percentage of subgiants obtained
via the $\log g$ versus $J-H$ diagram (18\%) and the one estimated by
using the Besan\c con Galaxy model (25\%) supports the values in $\log g$ of 
RAVE dataset. The percentage of evolved (subgiant and giant)
stars found in this work is consistent with \citet{Seabroke08} and in close
agreement with \citet{Klement11}. For the first time in the literature, we
analysed the effect of the CHISQ in the RAVE data.  Neither the CHISQ values
nor the signal-to-noise ratio bias RAVE $\log g$ values.  Therefore the method
of identifying dwarfs and giants via the two magnitude diagram has been verified
against an unbiased dataset.

\section{Acknowledgements}
We would like to thank the referee Dr. Antonio Cabrera-Lavers for
his useful comments that improved the readability of this paper.

Funding for RAVE has been provided by: the Australian Astronomical
Observatory, the Astrophysical Institute Potsdam, the Australian
National University, the Australian Research Council, the French
National Research Agency, the German Research Foundation, the Instituto
Nazionale di Astrofisica at Padova, The Johns Hopkins University, the
W.M. Keck Foundation, the Macquarie University, the Netherlands
Research School for Astronomy, the Natural Sciences and Engineering
Research Council of Canada, the Slovenian Research Agency, the Swiss
National Science Foundation, the Science \& Technology Facilities
Council of the UK, Opticon, Strasbourg Observatory, and the
Universities of Groningen, Heidelberg, and Sydney.

Salih Karaali is grateful to the Beykent University for financial
support. This publication makes use of data products from the Two
Micron All Sky Survey, which is a joint project of the University of
Massachusetts and the Infrared Processing and Analysis
Center/California Institute of Technology, funded by the National
Aeronautics and Space Administration and the National Science
Foundation.

This research has made use of the SIMBAD database, operated at CDS,
Strasbourg, France and NASA's Astrophysics Data System.


\begin{thebibliography}{99}

\bibitem[Ammons et al.(2006)]{Ammons06}
Ammons S. M., Robinson S. E., Strader J., Laughlin G., Fischer D., Wolf A., 2006,
ApJ, 638, 1004

\bibitem[Bahcall \& Soneira(1980)]{Bahcall80}
Bahcall J. N., Soneira R. M., 1980, ApJS, 44, 73

\bibitem[Bilir et al.(2006)]{Bilir06a}
Bilir S., Karaali S., G\"uver T., Karata\c{s} Y., Ak S., 2006, AN, 327, 72

\bibitem[Bilir et al.(2008)]{Bilir08}
Bilir S., Karaali S., Ak S., Yaz E., Cabrera-Lavers A., Co\c skuno\u glu K. B., 2008, MNRAS, 390, 1569

\bibitem[Co\c skuno\u glu et al.(2011)]{Coskunoglu11}
Co\c skuno\u glu B., et al., 2011, MNRAS, 412, 1237

\bibitem[Covey et al.(2007)]{Covey07}
Covey K. R., et al., 2007, AJ, 134, 2398

\bibitem[ESA(1997)]{ESA97}
ESA, 1997, Hipparcos and Tycho Catalogues, ESA SP-1200, 1, 57

\bibitem[Fouque et al.(2000)]{Fouque00}
Fouque P., et al., 2000, A\&AS, 141, 313

\bibitem[H{\o}g et al.(2000)]{Hog00}
H{\o}g E. E., Fabricius C., Makarov V. V., Urban S., Corbin T., Wycoff G.,
Bastian U., Schwekendiek P., Wicenec A., 2000, A\&A, 355, L27

\bibitem[Klement et al.(2011)]{Klement11}
Klement R. J., Bailer-Jones C. A. L., Fuchs B., Rix H.-W., Smith K. W., 2011,
ApJ, 726, 103

\bibitem[Marigo et al. (2008)]{Marigo08}
Marigo P., Girardi L., Bresson A., Groenewegen M. A. T., Silva L., Granato
G. L., 2008, A\&A, 482, 883

\bibitem[Marshall et al.(2006)]{Marshall06}
Marshall D. J., Robin A. C., Reyl\'{e} C., Schultheis M., Picaud S.,
2006, A\&A, 453, 635

\bibitem[Robin et al. (2003)]{Robin03}
Robin A. C., Reyl\'e C., Derri\`ere S., Picaud S., 2003, A\&A, 409, 523

\bibitem[Schlegel et al.(1998)Schlegel, Finkbeiner \& Davis]{Schlegel98}
Schlegel D. J., Finkbeiner D. P., Davis M., 1998, ApJ, 500, 525

\bibitem [Schultheis et al. (2006)]{Schultheis06}
Schultheis M., Robin A., C., Reyl\'e C., McCracken H. J., Bertin E.,
Mellier Y., Le F\'evre O., 2006, A\&A, 447, 185

\bibitem[Seabroke et al.(2008)]{Seabroke08}
Seabroke G. M., et al., 2008, MNRAS, 384, 11

\bibitem[Smith et al.(2007)]{Smith07}
Smith M., et al., 2007, MNRAS, 379, 755

\bibitem[Skrutskie et al.(2006)]{2mass06}
Skrutskie M. F., et al., 2006, AJ, 131, 1163

\bibitem[Steinmetz(2003)]{Steinmetz03}
Steinmetz M., 2003, ASP Conf. Ser., 298, 381

\bibitem[Steinmetz et al.(2006)]{Steinmetz06}
Steinmetz M., et al., 2006, AJ, 132, 1645

\bibitem[Veltz et al.(2008)]{Veltz08}
Veltz L., et al., 2008, A\&A, 480, 753

\bibitem[Yanny et al.(2009)]{Yanny09}
Yanny B., et al., 2009, AJ, 137, 4377

\bibitem[York et al.(2000)]{York00}
York D. G., et al., 2000, AJ, 120, 1579

\bibitem[Zwitter et al.(2008)]{Zwitter08}
Zwitter T., et al., 2008, AJ, 136, 421

\bibitem[Zwitter et al.(2010)]{Zwitter10}
Zwitter T., et al., 2010, A\&A, 522A, 54

\end{thebibliography}
\end{document}